# An Ontology-based Context Model in Intelligent Environments

Tao Gu [1,2], Xiao Hang Wang [1,2], Hung Keng Pung [1], Da Qing Zhang [2]

[1] Department of Computer Science, National University of Singapore, Singapore

[2] Connected Home Lab, Institute for Infocomm Research, Singapore

gutao@comp.nus.edu.sg, xwang@i2r.a-star.edu.sg,
punghk@comp.nus.edu.sg, daqing@i2r.a-star.edu.sg

**Abstract:** Computing becomes increasingly mobile and pervasive today; these changes imply that applications and services must be aware of and adapt to their changing contexts in highly dynamic environments. Today, building context-aware systems is a complex task due to lack of an appropriate infrastructure support in intelligent environments. A context-aware infrastructure requires an appropriate context model to represent, manipulate and access context information. In this paper, we propose a formal context model based on ontology using OWL to address issues including semantic context representation, context reasoning and knowledge sharing, context classification, context dependency and quality of context. The main benefit of this model is the ability to reason about various contexts. Based on our context model, we also present a Service-Oriented Context-Aware Middleware (SOCAM) architecture for building of context-aware services.

## 1 INTRODUCTION

The advanced deployment of wireless networks and mobile devices is moving computing towards a new field knows as pervasive computing in which devices and services are seamlessly cooperated to support users' tasks. Emerging pervasive computing technologies provide "anytime, anywhere" computing by decoupling users from devices and viewing applications as entities that perform tasks on behalf of users [1]. To avoid increasing complexity, and allow the user to concentrate on his tasks, applications must be capable to operate in highly dynamic environments. Devices, services and agents in pervasive computing environments must be aware of their contexts and automatically adapt to their changing contexts - know as context-awareness. By context, we refer to any information that can be used to characterize the situation of an entity, where an entity can be a person, place, or physical or computational object [2].

Context-aware computing has been drawing much attention from researchers since it was proposed about a decade ago. A number of context-aware systems have been developed to demonstrate the usefulness of this new technology, such as Context Toolkit [3], HP's Cooltown [4] and MIT's AIRE spaces [5], whereas some other systems are still under research, such as Context Fabric [6], CoBrA [7] and GAIA [8]. However, context-aware services have never been widely available to everyday users. Building context-aware systems is still a complex and time-consuming task due to lack of an appropriate infrastructure or middleware-level support. An appropriate infrastructure for context-aware systems should provide support for most of the tasks involved in dealing with contexts - acquiring context from various sources such as physical sensors, databases and agents; performing context interpretation; carrying out dissemination of context to interested parties in a distributed and timely fashion; and providing programming models for constructing of context-aware services. To support these tasks, a context model needs to be well established.

In this paper, we present a context model based on ontology using OWL - Web Ontology Language [9] to support various tasks in our context-aware middleware. It supports semantic context representation by defining the common upper ontology for context information in general; and providing a set of low-level ontologies which apply to different sub-domains. It models the basic concepts of person, location, computational entity and activity; describes the properties and relationships between these concepts. Our context model captures various contexts by introducing a classification scheme; captures relationships between different context information by introducing dependency tag to the property associated with a specified context class; captures quality of context by annotating sensed context with extensible quality constraints. It also supports the use of different context reasoning engines to reason about various contexts so that applications can be given a notion of the confidence of different contexts before acting on it. In this paper, we also present a Service-Oriented Context-Aware Middleware (SOCAM) architecture for the building and rapid prototyping of context-aware services in intelligent environments.

The rest of this paper is organized as follows. Section 2 begins the discussion on related work. In section 3 we review and discuss the OWL language. In Section 4 we describe our modeling concept, followed by the architecture design in Section 5. Finally, we conclude in section 6.

## 2 RELATED WORK

Much research has been done in the area of context-aware computing in the past few years. In this section, we review and discuss some important context models. We classify the existing context models into three categories:

*Application-oriented approach*: Many existing context-aware systems model and represent context only for specific applications. These models typically are proprietary and exploratory, and lack formality and expressiveness. The HP's Cooltown project proposed a web-based context model in which each object (person, place and thing) has a corresponding web description that can be retrieved using a URL. The Context Toolkit project transmits low-level context acquired from physical sensors to the form of XML-encoded name-value pairs.

*Model-oriented approach*: This category of models commonly uses conceptual modeling approaches to represent context. A formal context model based on ER model was proposed by several projects [10][11]; and context can be easily managed with relational databases. Henricksen et al. [12] model contexts and their additional features (classification and temporal characteristics) using both ER model and UML diagrams. This model was further reformulated with the extended Object-Role Modeling (ORM) [13].

*Ontology-oriented approach*: Some work in the field of context-awareness ignore issues about quantitative concepts including temporal characteristics and quality of context, and focused more on constructing an ontology for context in a specific domain to reach the goals of knowledge sharing across distributed systems. The Comprehensive Structured Context Profiles (CSCP) [14] was developed based on RDF to represent context by means of session profiles. Chen et al. defined a context ontology based on OWL to support ubiquitous agents in their Context Broker Architecture (CoBrA), this context ontology only covers contexts in campus space, while has no explicit support for modeling general contexts in heterogeneous environments. Ranganathan et al. [8] developed a middleware for context awareness and semantic interoperability, in which they represented context ontology using DAML+OIL [15].

Of the above three categories, the application-oriented approach lacks the formal basis and does not support knowledge sharing across different systems. Though the model-oriented projects support formality and some of them capture temporal aspect of context information, they do not address issues including knowledge sharing and context reasoning. The ontology-oriented approach focuses on context ontology and explores the potential capability of context reasoning based on Semantic Web technologies. However, the existing context ontologies lack of generality and have not addressed important issues including context classification, context dependency and quality of context which will be useful in context reasoning. In this paper, we present our ontology-based context model using OWL that addresses these shortcomings.

## 3 OWL

OWL is a language for defining ontologies. Ontology is referred as the shared understanding of some domains, which is often conceived as a set of entities, relations, functions, axioms and instances.

We have chosen OWL to realize our context model and define our context ontologies for three reasons. First, it is much expressive compared to other ontology languages such as RDFS [16]. Second, it has the capability of supporting semantic interoperability to exchange and share context knowledge between different systems, i.e., contexts can be exchanged and understood between different systems in various domains; and enabling automated reasoning to be used by automated processes. Last, we chose OWL rather than DAML+OIL as DAML+OIL is merging into OWL to become an open W3C standard.

## 4 AN ONTOLOGY-BASED MODEL

In this section, we will describe our design considerations and modeling concepts, together with a context-aware home scenario to be used to illustrate our context model.

### 4.1 A Context-Aware Home Scenario

A context-aware home is a smart home environment which is equipped with various networked sensors/actuator devices such as cameras, microphones, RFID (Radio Frequency Identification) based location sensors, X.10 curtain sensors, etc. In this section, we describe a typical scenario in order to illustrate our modeling concept.

Daddy John carrying a cell phone has entered his house; the face recognition system senses his presence and his location information get updated. When John moves into the bathroom to take a shower or goes to his bedroom for a nap, his personal communication agent interprets his current status by using the contexts acquired from various sensors and decides to forward all phone calls to his voice mail box.

Mom Julia comes back from shopping with her baby girl and her 5-years-old son -Tom. She settles down her baby in the baby's room, and then enters to the kitchen. An audio/visual communication channel can be established between the kitchen and her baby's room. When she moves around the rooms, the communication channel is able to automatically switch and remain alive between Julia and her baby. Thus, Julia is able to have a face-to-face talk with her baby using the embedded video conferencing panel in each room just like she is in her baby's room.

Julia wants to have a Barbeque dinner outside the house tonight. She quickly consults her meal arrangement agent which is able to advise her whether it is possible. The meal arrangement agent consults the networked fridge for available food items based on their food preferences of all family members and queries on an external weather service for the weather condition tonight. After a while, she realizes that the Barbeque dinner is not possible due to weather condition. After dinner, when Julia sits on the sofa in the living room and turns on the TV, the lighting begins to dim.

## 4.2 Design Considerations

A context-aware system requires context information to be exchanged and used between different entities such as users, devices and services in a same semantic understanding. In other word, an appropriate context model should support semantic interoperability which enables the common schemas to be shared between different entities. For example, in the above scenario, the representation of John's location should be understood between his personal communication agent and his cell phone.

Context information exhibits a number of characteristics in intelligent environments. First, context information has a great variety. The definition of context includes any information that describes physical objects, applications and users in any domain. Second, context information varies in different sub-domains. For example, we are more concerned about device context such as fridge, TV and DVD player in a home environment whereas workstation and PC in an office environment. Third, context information is interrelated. For example, in our scenario, Julia's current status (watching TV) are closely related to where she is located (located at LivingRoom), where the TV is located (located at LivingRoom), and her TV's current status (ON). Fourth, context information is inconsistent. For example, in our scenario, Tom's location context may quickly become out-of-date when he is rushing into different rooms. Physical sensors may also cause context conflict, for example, the bedroom location sensor may sense Tom is not present in his bedroom whereas the camera senses his presence.

## 4.3 Context Ontology

The basic concept of our context model is based on ontology which provides a vocabulary for representing knowledge about a domain and for describing specific situations in a domain. Context ontology defines a common vocabulary to share context information in a pervasive computing domain; and include machine-interpretable definitions of basic concepts in the domain and relations among them. The main advantage of our context model is sharing common understanding of the structure of context information among users, devices and services to enable semantic interoperability. It also enables reuse of domain knowledge, i.e., building a large ontology by integrating several ontologies describing portions of the large domain. Most importantly, it enables formal analysis of domain knowledge, for example, context reasoning becomes possible by explicitly defining context ontology.

The context ontology should be able to capture all the characteristics of context information. First, it is responsible to capture a great variety of context. To capture various contexts in a pervasive computing environment is indeed a difficult task which many researchers face. As the pervasive computing domain can be divided into a collection of sub-domains such as home domain, office domain, vehicle domain, open space domain, etc, it would be easy to specify the context in one domain in which a specific range of context is of interest. The separation of domain can also reduce the burden of context processing and make it possible to interpret context information on mobile thin clients. Our context ontologies are divided into upper ontology and domain-specific ontologies. The upper ontology is a high-level ontology which captures general context knowledge about the physical world in pervasive computing environments. The domain-specific ontologies are a collection of low-level ontologies which define the details of general concepts and their properties in each sub-domain. The low-level ontology in each sub-domain can be dynamically plugged into and unplugged from the upper ontology when the environment is changed, for example, when a user leaves his home to drive a car, the home-domain ontology can be automatically unplugged from the system; and the vehicle-domain ontology can be plugged into the system.

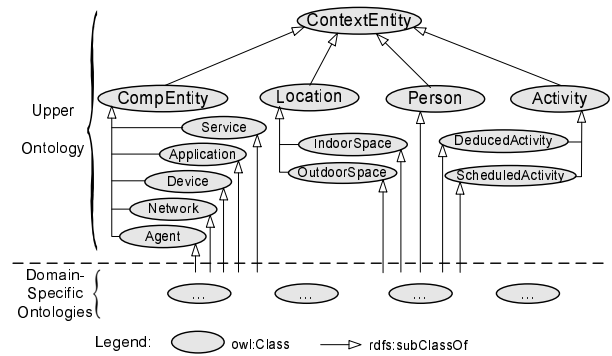

**Figure 1.** Class hierarchy diagram for our context ontologies

The upper ontology defines the basic concepts of person, location, computational entity and activity as shown in Figure 1. The class *ContextEntity* provides an entry point of reference for declaring the upper ontology. One instance of *ContextEntity* exists for each distinct user, agent or service. Each instance of *ContextEntity* presents a set of descendant classes of *Person*, *Location*, *CompEntity* and *Activity*. The details of these basic concepts are defined in the domain-specific ontologies which may vary from one domain to another. We have defined all the descendant classes of these basic classes in a smart home environment and a set of properties and relationships that are associated with these classes.

### 4.4 Modeling Classification and Dependency

We classify a wide range of contexts into two main categories - direct context and indirect context based on the means by which context is obtained. Direct context is acquired from a context provider directly. A context provider can be an internal source such as an indoor location provider, or an external source such as a weather information server. Direct context can be further classified into sensed context and defined context. Sensed context is obtained from physical sensors, for example, curtain's status context sensed by curtain sensors, or from virtual sensors, for example, a web service. Defined context is typically defined by a user. They may have different invariant periods from days to years, for example a person's name - "John" and his date of birth are invariant over its lifetime whereas John's food preference may be changed over a couple of months.

Indirect context is obtained by interpreting direct context through aggregation and reasoning process. By aggregating direct context, for example, John's food preference and Julia's food preference in our scenario, we can obtain aggregated context such as Smith family's food preferences. By using context reasoning engine, deduced context can be obtained and inferred from other types of context, for example, John's current status context (Sleeping) is inferred from his location context (MasterBedroom), his posture context (LiedDown), the curtain's status context (NotOpen) and the door's status context (Close).

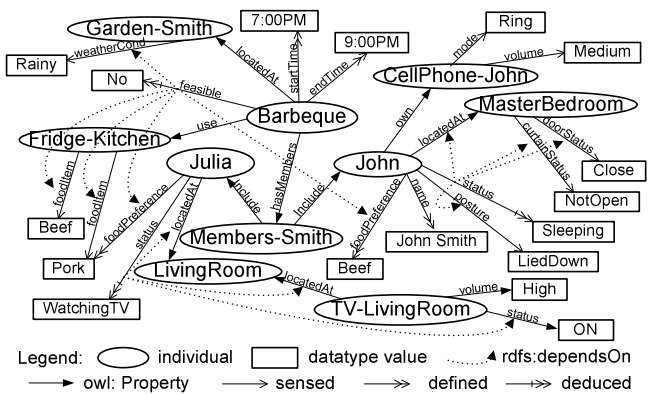

**Figure 2.** A partial OWL/RDF graph notation for interreated contexts in the scenario of Section 4.1. (John and Julia is type of the class Person, LivingRoom and MasterBedroom is type of the class IndoorSpace, Garden-Smith is type of the class OutdoorSpace, Barbeque is type of the class ScheduledActivity, CellPhone-John and Fridge-Kitchen is type of the class Device, Members-Smith is type of the class FamilyMember.)

By introducing context classification information in our context model, we are able to perform context reasoning based on confidence level of each type of context as we will illustrate in Section 4.6. We present a graph representation of our context model based on the scenario described in Section 4.1 as shown in Figure 2.

To describe context classification information in our context ontologies using OWL, we introduce an additional property element - *owl:classifiedAs* in the property restriction. This special element is able to capture the properties of context classification associated with datatypes and objects. In our context ontologies, this additional property will have the values such as Sensed, Defined, Aggregated or Deduced. Figure 3 shows an example of describing classification information - Defined in the ObjectProperty - *hasChildren*.

```
<owl:Class rdf:ID="Person">
    <rdfs:subClassOf>
        <owl:Restriction>
            <owl:onProperty rdf:resource="hasChildren"/>
            <owl:to Class rdf:resource="#Person"/>
            <owl:classifiedAs rdfs:resource="http://lucan.ddns.comp.nus.edu.sg/octopus/classification#Defined"/>
        </owl:Resrtiction>
    </rdfs:subClassOf>
</owl:Class>
```

**Figure 3.** An OWL expression for describing classification information

Dependency is an important characteristic of context information as we pointed out in section 4.2. A dependency captures the existence of a reliance of property associated with one entity on another. For example, in our scenario, Julia's current status depends on where she is located (*locatedAt*), where the TV is located (*locatedAt*) and the TV's current status (*status*). To describe dependency information using OWL, we introduce an additional property elements -*rdfs:dependsOn* in both object property and data property. This special element is able to capture the dependency relationship of properties associated with datatypes and objects. The example in Figure 4 shows the feasible property of *ScheduledActivity* class depends on where the person is located (*locatedAt*), weather condition (*weatherCond*), etc.

```
<owl:ObjectProperty rdf:ID="feasible">
    <rdfs:domain rdf:resource="ScheduledActivity"/>
    <rdfs:classifiedAs rdfs:resource="http://lucan.ddns.comp.nus.edu.sg/octopus/classification#Defined"/>
    <rdfs:dependsOn rdf:resource="locatedAt"/>
    <rdfs:dependsOn rdf:resource="weatherCond"/>
    ...
</owl:ObjectProperty>
```

**Figure 4.** An OWL expression for describing dependency information

### 4.5 Modeling Quality of Context

Context information is inconsistent due to highly dynamic nature of pervasive computing systems and imperfect sensing technology. The location context may vary in a few seconds when a person moves around the rooms. Physical sensors may produce incorrect or stale context data due to poor reliability and processing delay of converting low-level sensor data to high-level context.

Our context ontology allows the properties of entities to be associated with quality constraints that indicated the quality of context. We have constructed an extensible ontology for quality of information. As shown in Figure 5(a), Quality Constraints are used as quality indicators of OWL properties. Quality Constraints are associated with a number of quality parameters, which capture the dimensions of quality relevant to the attributes of entities and relationships between entities. Each parameter is described by one or more appropriate quality metrics, which defines how to measure or compute context quality with respect to the parameter. Besides a value, a metric contains a type and a unit.

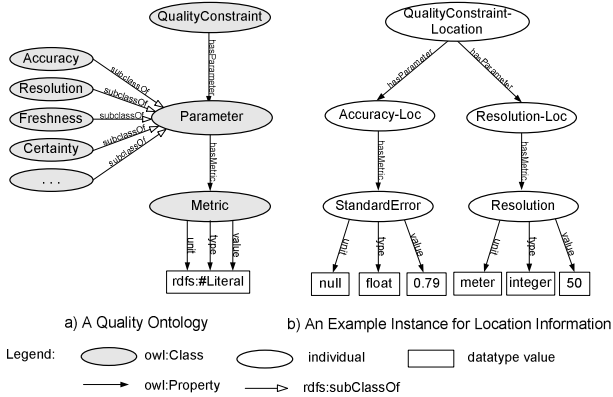

**Figure 5.** Ontology and an example instant for Quality Constraint

We have defined four types of quality parameters that are most commonly used: accuracy - range in terms of a measurement; resolution - smallest perceivable element; certainty - the probability to describe the state of being certain and freshness - production time and average lifetime of a measurement. The example in Figure 5(b) illustrates the use of quality constraint to define the quality information about a person's location - a piece of sensed context which provides location information in terms of coordinates with a resolution of 50 meters and an accuracy of 79%.

### 4.6 Context Reasoning

The important feature of our context model is the ability to support automated context reasoning which is the process of reasoning about various types of contexts and their properties. Context reasoning broadens context information implicitly by introducing deduced context derived from other types of context. It also provides a solution to resolve context inconsistency and conflict that caused by imperfect sensing.

By reasoning context, deduced context can be inferred from sensed, defined or aggregated context based on our context classification scheme. For example, in our scenario, Deduced context (John's current status) can be inferred from sensed context (John's location and posture, Door's and Window's status) as illustrated below using first-order logic predicates.

*Location(John, MasterBedRoom)* ∧ *Posture(John, LiedDown)* ∧ *Status(Door, Close)* ∧ *( ¬ Status(Curtain, Open))* ├ *Status(John, Sleeping)*

A more complicated example below shows deduced context (Barbeque is not feasible) can be inferred from sensed context (Rainy, Fridge's food items), defined context (John's food preference) and aggregated context (All family members' food preferences).

*WeatherCond(Weather, Rainy)* ∧ *FoodPreference(Members, FoodItems)* ∧ *Available(Fridge, FoodItems)* ├ *Feasible(Barbeque, NO)*

By reasoning context classification information and quality information based on our context model, we are able to detect and resolve context conflict. Different types of context have different levels of confidence and reliability, for example, defined context is more reliable compared to sensed and deduced context; and also have different levels of quality, for example, a RFID-based location sensor may have a 80% accuracy rate whereas a Bluetooth-based location sensor may only have a 60% accuracy rate.

## 5 ARCHITECTURE OVERVIEW

In this section, we describe our service-oriented context-aware middleware (SOCAM) architecture. Our architecture aims to help application programmers to build context-aware services more efficiently. The SOCAM architecture consists of the following components as shown in Figure 6:

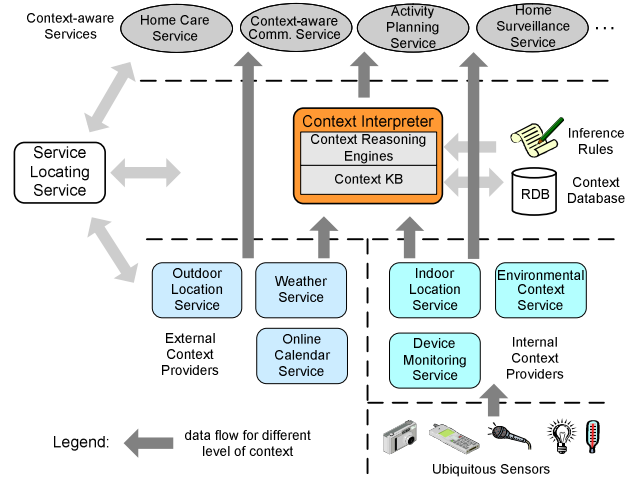

**Figure 6.** Overview of the SOCAM architecture

*Context Providers*: Context Providers abstract contexts from different sources - External Context Providers or Internal Context Providers; and convert them to OWL representation so that contexts can be shared and reused by other SOCAM components.

*Context Interpreter*: Context Interpreter consists of Context Reasoning Engines and Context KB (Knowledge Base). The Context Reasoning Engines provide the context reasoning services including inferring deduced contexts, resolving context conflicts and maintaining the consistency of Context KB. Different inference rules can be specified and input into the reasoning engines. The context KB provides the service that other components can query, add, delete or modify context knowledge stored in the Context Database.

*Context-aware Services*: Context-aware Services make use of different level of contexts and adapt the way they behave according to the current context.

*Service Locating Service*: Service Locating Service provides a mechanism where the Context Providers and the Context Interpreter can advertise their presences; users or applications can locate and access these services.

Based on the SOCAM architecture, we currently implementing a prototype system that aims to realize the context-aware home scenario that we have described in Section 4.1. It consists of an OSGi-compliant residential gateway which connects the home network to the Internet; and various computing devices and physical sensors in a smart home environment. The Context Interpreter will be running on the OSGi gateway and implemented based on HP's semantic web toolkit - Jena2 [17]. The Service Locating Service has been developed in our service discovery project [18].

# 6 CONCLUSION

In this paper, we have presented a formal and extensible context model based on OWL to represent, manipulate and access context information in intelligent environments. Our context model represents contexts and their classification, dependency and quality information using OWL to support semantic interoperation, context knowledge sharing and context reasoning. We are looking at different context reasoning mechanisms for reasoning about various contexts such as first-order probabilistic logic, high-order logic and Bayesian networks. We will also continue our work on building a prototype system in a smart home environment.